\documentclass[conference]{IEEEtran}

\usepackage{cite}
\usepackage{amsmath,amssymb,amsfonts}
\usepackage{algorithmic}
\usepackage{graphicx}
\usepackage{textcomp}
\usepackage{xcolor}
\usepackage{makecell}
\usepackage{tcolorbox}
\usepackage[table]{xcolor}
\usepackage{multirow}
\usepackage{geometry}
\usepackage{capt-of} % Per \captionof
\usepackage{flushend}

\usepackage{mdframed}

\def\BibTeX{{\rm B\kern-.05em{\sc i\kern-.025em b}\kern-.08em
    T\kern-.1667em\lower.7ex\hbox{E}\kern-.125emX}}
\begin{document}

\newcommand{\biagio}[1]{\textcolor{red}{Biagio: #1}}

\title{Prompt Engineering vs. Fine-Tuning for LLM-Based Vulnerability Detection in Solana and Algorand Smart Contracts}

\author{\IEEEauthorblockN{Biagio Boi}
\IEEEauthorblockA{\textit{dept. of Computer Science} \\
\textit{University of Salerno}\\
Fisciano, Salerno, Italy \\
bboi@unisa.it}
\and
\IEEEauthorblockN{Christian Esposito}
\IEEEauthorblockA{\textit{dept. of Computer Science} \\
\textit{University of Salerno}\\
Fisciano, Salerno, Italy \\
esposito@unisa.it}
}

\maketitle

\begin{abstract}
Smart contracts have emerged as key components within decentralized environments, enabling the automation of transactions through self-executing programs. While these innovations offer significant advantages, they also present potential drawbacks if the smart contract code is not carefully designed and implemented. This paper investigates the capability of large language models (LLMs) to detect OWASP-inspired vulnerabilities in smart contracts beyond the Ethereum Virtual Machine (EVM) ecosystem, focusing specifically on Solana and Algorand. Given the lack of labeled datasets for non-EVM platforms, we design a synthetic dataset of annotated smart contract snippets in Rust (for Solana) and PyTeal (for Algorand), structured around a vulnerability taxonomy derived from OWASP. We evaluate LLMs under three configurations: prompt engineering, fine-tuning, and a hybrid of both, comparing their performance on different vulnerability categories. Experimental results show that prompt engineering achieves general robustness, while fine-tuning improves precision and recall on less semantically rich languages such as TEAL. Additionally, we analyze how the architectural differences of Solana and Algorand influence the manifestation and detectability of vulnerabilities, offering platform-specific mappings that highlight limitations in existing security tooling. Our findings suggest that LLM-based approaches are viable for static vulnerability detection in smart contracts, provided domain-specific data and categorization are integrated into training pipelines.
\end{abstract}

\begin{IEEEkeywords}
Solana, Algorand, Vulnerability Assessment, Smart Contract, LLM
\end{IEEEkeywords}

\section{Introduction}

Smart contracts represent the backbone of decentralized applications built on top of blockchain platforms, enabling the automation of transactions through self-executing programs. While these innovations offer significant advantages—such as increased transparency in transactions, which has made a substantial impact on areas like supply chain management \cite{cai2023impact} and digital certification systems \cite{hammoudeh2023digital}—they also present potential drawbacks if the smart contract code is not carefully designed and implemented. These self-executing programs that reside on the blockchain are particularly exposed due to their public visibility, permanent deployment, and financially critical logic. Once deployed, these contracts cannot be altered, making any flaws in their design or implementation particularly dangerous. 

Large Language Models (LLMs) are increasingly used to generate code, including smart contracts for blockchain platforms like Ethereum, with an increasing delegation of code generation to such tools \cite{napoli2024leveraging}. LLMs are also being used for the verification of the code written by programmers, as they are able to find bugs and vulnerabilities there. 

LLMs have proved themselves to be useful for code generation, while vulnerability assessment is still at its beginning with a not satisfying prediction \cite{10456393}. In addition, when users prompt LLMs to write smart contracts, the model generates code based on its training data. However, the model might rely on incomplete knowledge about security standards, and, more critically, they have not been properly fine-tuned for verification compared to the mere syntactic check. Given the recent expansion of alternative blockchain platforms, each characterized by distinct programming languages and architectural frameworks \cite{bartoletti2024smart}, the role of vulnerability assessment is progressively more important than in the past. On the one hand, the current literature lacks a proper analysis of the vulnerability types outside the Solidity and Ethereum framework. On the other hand, despite having tools available for blockchains such as Solana and Algorand, which differ significantly from Ethereum, there is no precise evaluation of their effectiveness in detecting the most common vulnerabilities.
In light of current needs and to overcome these limitations, in this study, we:
\begin{itemize}
\item Investigate security challenges across emerging blockchain platforms—specifically Algorand and Solana—by analyzing critical vulnerabilities according to the recent OWASP Top 10 \cite{owasp_top10_2025}.

\item Examine the role of LLMs in vulnerability detection for domain-specific programming languages, and compare the effectiveness of prompt engineering and fine-tuning techniques.
\end{itemize}
%The manuscript is organized into six sections. Section 2 gives a background on vulnerability assessment and LLM-effectiveness strategies. Section 3 provides a comprehensive review of the current state of the art in LLM-based vulnerability assessment. It identifies key gaps in the literature and formulates three research questions (RQs) to guide the study according to the outlined objectives.  Section 4 outlines the methodology to address the RQs. Section 5 reports the experimental results and evaluates the proposed approach by benchmarking it against existing tools, with a particular focus on performance comparison involving two state-of-the-art LLMs. Finally, Section 6 concludes the paper and discusses promising directions for future research in the area of LLM-driven vulnerability detection.
\section{Background}
\subsection{Blockchain Security}
The deterministic execution of smart contracts requires all nodes in the network to arrive at the same state, meaning that even a single exploitable bug can be globally catastrophic. While traditional security models such as CIA (Confidentiality, Integrity, Availability) remain partially applicable, the decentralized and immutable nature of blockchains introduces new attack surfaces. Most popular vulnerabilities in smart contracts can be categorized into various types. (1) Control flow issues, such as reentrancy attacks, where recursive function calls allow an attacker to drain funds before state updates are completed. (2) Arithmetic errors, including integer underflow and overflow, which can alter logic or bypass checks. (3) Access control failures, where insecure permission structures grant unintended users the ability to modify critical variables. (4) Unchecked external calls, which can lead to unexpected interactions with untrusted code.

Recent work (e.g., OWASP Top 10 for Blockchain, 2025 \cite{owasp_top10_2025}) categorizes these threats into protocol-level attacks, implementation-level bugs, and vulnerabilities within smart contract logic. Several widely adopted taxonomies have emerged to describe and organize these risks. The most relevant one, apart OWASP Top 10, is the SWC Registry (Smart Contract Weakness Classification and Test Cases), a structured vulnerability repository maintained by the smart contract auditing community. 
Anyway, to effectively assess vulnerabilities across different blockchain platforms requires dedicated evaluation tailored to their execution environments and programming models. For instance, Ethereum’s EVM supports tools like SmartCheck \cite{tikhomirov2018smartcheck} and Slither\cite{feist2019slither}, but those cannot be applied to Algorand and Solana blockchains due to different characteristics both in programming language and blockchain structure. As an example, while Algorand leverages TEAL programming language, Solana leverages Rust. Another example is that aforementioned Ethereum-specific tools often rely on symbolic execution, data flow analysis, or fuzzing techniques to surface vulnerabilities prior to deployment, which requires knowing the environment in which the execution will take place. To better explain these differences, in this section, we briefly introduce some concepts about the characteristics of Algorand and Solana blockchains. In the subsequent section, we delve deeper into the specific vulnerabilities affecting these platforms and analyze how the LLM can manage these vulnerabilities.

\subsubsection{Algorand}
The Algorand blockchain was born to overcome a critical aspect of decentralized systems, known as \emph{Blockchain Trilemma}, characterized by a tradeoff of decentralization, security, and scalability. It is hard to maintain all the three properties in a blockchain. According to a recent evaluation presented in \cite{10740107}, Algorand it is able to easily surpass the Ethereum 2.0 performance by offering increased indicators for all three characteristics. Unlike traditional blockchains that rely on energy-intensive proof-of-work mechanisms, Algorand uses a Pure Proof-of-Stake (PPoS) consensus algorithm. This allows it to process thousands of transactions per second with finality in seconds. Unlike permissioned blockchains or those with limited validator sets, Algorand’s PPoS protocol randomly selects participants from the entire set of token holders. This randomness and inclusiveness make it resistant to centralization and censorship. Considering the strong consensus protocol, most of the vulnerabilities come from the smart contract logic,  transaction configuration, and application-layer design ~\cite{sun2023panda}. During the validation of the tool discussed in ~\cite{sun2023panda}, it was observed that approximately 4.04\% of on-chain applications exhibited vulnerabilities, with the majority of these issues arising from arbitrary deletion operations. As will be elaborated in subsequent sections of this manuscript, vulnerabilities can be easily introduced by smart contract programmer, despite having clear in mind which are the main drawbacks they can face.

\subsubsection{Solana}
Solana is a permissionless blockchain designed for scalable decentralized applications. Differently from Ethereum’s account-based model, Solana employs a parallel execution model known as Sealevel, allowing non-conflicting smart contracts (called programs) to run concurrently across multiple cores, and increasing its Transaction-per-second (TPS). At its core, Solana introduces a unique consensus innovation called Proof of History (PoH), which acts as a cryptographic timestamping mechanism. PoH enables the network to order transactions efficiently without relying entirely on traditional consensus messages. This is combined with Tower BFT, a practical Byzantine Fault Tolerant (PBFT) algorithm, to finalize blocks. Solana achieves high throughput—reportedly over 65,000 transactions per second—with low latency and low fees. Smart contracts are written in Rust and deployed as BPF bytecode. Solana’s Cross-Program Invocation (CPI) system allows on-chain programs to call each other, similar to Ethereum’s message calls, but with different security trade-offs \cite{wu2025exploring}. However, this architecture also introduces unique challenges for static analysis, particularly due to the program-centric design and explicit account management model.

\subsection{LLMs for Vulnerability Assessment}
Smart contracts, although written in programming languages such as Solidity, Rust, or PyTeal, often encode logical structures and patterns that are linguistically interpretable. LLMs trained on code—particularly those fine-tuned on security-focused context can reason about these patterns in a manner similar to natural language processing. This capability enables them to predict insecure code snippets based on known vulnerability templates and to provide natural language explanations or risk assessments. This dual ability positions LLMs as viable tools for vulnerability assessment in blockchain-based applications.

\subsubsection{Fine-Tuning}
Fine-tuning refers to the process of adapting a pre-trained LLM to a specific domain or task by continuing its training on a curated dataset. In the context of smart contract security, this involves exposing the model to a dataset composed of smart contract code annotated with vulnerabilities—both in terms of their presence and their categorization. The dataset typically consists of input–output pairs, where each input is a code snippet and the output is a vulnerability label or description. During the training phase, the model adjusts its internal parameters to better capture the semantics of these vulnerability patterns. Once training is complete, the model enters an inference phase, where its performance can be validated on previously unseen data to assess the effectiveness of the fine-tuning process. Despite its advantages, fine-tuning introduces challenges such as overfitting (especially when the dataset is small), loss of generalizability to other domains, and high computational costs. Moreover, there is a risk that fine-tuning on a narrow domain could overwrite useful general-purpose knowledge, which may not be desirable in some applications.

\subsubsection{Prompt Engineering}
Prompt engineering has emerged as a lightweight and adaptable method to guide LLM behavior without modifying the underlying model architecture or weights. Common strategies include:

\begin{itemize}
    \item Zero-shot prompting: Asking the model to detect vulnerabilities without providing any prior examples.
    \item Few-shot prompting: Supplying the model with a few examples of vulnerable code snippets and their explanations to improve its understanding of the task.
    \item Role-based prompting: Instructing the model to "act as a security auditor" or "analyze the following smart contract like a formal verifier" to condition its response behavior.
\end{itemize}

These techniques influence the model's capacity to detect both syntactic issues (e.g., uninitialized variables, missing checks) and semantic vulnerabilities (e.g., logic flaws, improper access control). While fine-tuning offers a more permanent and domain-specific adaptation of the model, it also introduces rigidity and potential knowledge degradation. In contrast, prompt engineering retains the flexibility of the base model, making it a suitable alternative when the LLM already exhibits reasonable competence in code understanding. Depending on the application context, a hybrid strategy—combining prompt engineering with selective fine-tuning—may offer the best trade-off between adaptability and performance.

\section{State of the Art}
Significant efforts have been dedicated to get insights into vulnerabilities in decentralized environments, particularly in the context of smart contracts \cite{bhardwaj2021attack, rautenberg2022case}, with a specific focus on Ethereum \cite{poston2020mapping}. The detailed analyses of Ethereum security have set a precedent for blockchain vulnerability assessment, and it is currently under investigation with a novel 2025 OWASP Top 10 release \cite{owasp_top10_2025}. As blockchain technologies' rapid evolution and diversification demand a more holistic approach, the necessity of extending these security assessments across a broader range of blockchain systems becomes increasingly apparent \cite{estrada2023systematic}. In response to this growing need, De Angelis et al. \cite{de2022evaluating} extensively evaluated the security properties of several blockchain platforms, including Ethereum, Algorand, and Hyperledger Fabric. Their findings highlight key security attributes, such as integrity, availability, accountability, and authorization, by revealing significant parallels between Ethereum and Algorand. This comparative analysis underscores the importance of examining different platforms and their merits while identifying commonalities that can inform cross-platform security practices. In contrast, other blockchain ecosystems, such as Solana, exhibit critical weaknesses, as determined in a subsequent study \cite{deangelis2023}, which underscores the variability in security postures between different blockchains. Moreover, the security-by-design principles that underpin Algorand as one of the most effective offer a level of architectural resilience that sets these platforms apart from their peers. To this end, we aim to investigate the role of vulnerabilities in two expanding blockchains: Algorand and Solana.

\begin{mdframed}[backgroundcolor=gray!10, linecolor=gray!80, linewidth=1pt] \textit{\textbf{RQ1}: How can the OWASP Top 10 EVM Vulnerabilities be mapped in Algorand and Solana blockchains?} \end{mdframed}

In parallel with statistical tool developments, the role of LLMs in the blockchain domain is gaining traction. LLMs are not only facilitating the automated generation of smart contracts from natural language specifications, as illustrated by Bartoletti et al. \cite{bartoletti2024smart}, and from high-level definition to low-level one such as Move Language \cite{karanjai2024solmover} but also expanding the possibilities for vulnerability detection and audit processes. Several recent studies have demonstrated the efficacy of LLMs in producing comprehensive audit reports and identifying vulnerabilities across multiple programming languages \cite{mothukuri2024llmsmartsec}. This multi-lingual capability, supported by ensemble learning approaches \cite{10664408}, enhances the model's ability to detect logic-based vulnerabilities, such as reentrancy \cite{boi2024smart, soud2024soley}, with a high degree of accuracy.

Despite these advancements, there remains a critical research gap in applying LLMs to other blockchain platforms beyond Ethereum. The security vulnerabilities of platforms such as Solana, which employs the Rust programming language, and Algorand, which employs the TEAL programming language, have yet to be rigorously explored in the context of LLM-driven vulnerability assessment. This gap limits the generalizability of current models and underscores the need for foundational research in expanding LLM capabilities to a wider array of blockchain ecosystems. 

\begin{mdframed}[backgroundcolor=gray!10, linecolor=gray!80, linewidth=1pt] \textit{\textbf{RQ2}: How effective are LLMs in detecting vulnerabilities in less-studied blockchain programming languages, such as Rust (Solana) and PyTeal (Algorand)?} \end{mdframed}

The introduction of prompt engineering techniques has significantly enhanced the performance of Large Language Models (LLMs) in vulnerability assessment tasks \cite{radcliffe2024automated, 10698605}. These techniques are highly dependent on the architecture and training scope of the model in use, and their effectiveness can vary accordingly. Among the most relevant approaches, we focus on zero-shot and few-shot prompting. The term "shot" refers to the number of example input-output pairs provided to the model before executing the target task. A comparative analysis presented in \cite{electronics13132657} highlights that few-shot techniques often outperform zero-shot in terms of accuracy for certain models. Moreover, strategies such as code injection and role-based prompting have been employed to further boost the reliability of LLMs, typically by guiding the model to perform classification or detection tasks more precisely \cite{10479384}.  While prompt engineering provides a lightweight way to adapt LLMs for vulnerability analysis, fine-tuning offers a more robust and specialized alternative by training models on blockchain-specific datasets. Both approaches have shown promising results in general-purpose code analysis, and in Ethereum blockchains \cite{ortu4530467identifying, 10679877}, however there is a notable lack of research specifically applying prompt engineering techniques to blockchain-based vulnerability assessments—particularly in underrepresented ecosystems such as Algorand and Solana, which use domain-specific languages (TEAL and Rust, respectively).

\begin{mdframed}[backgroundcolor=gray!10, linecolor=gray!80, linewidth=1pt]
\textit{\textbf{RQ3}: To what extent can prompt engineering techniques (role-based prompting) and model fine-tuning improve the accuracy of LLM-based vulnerability assessments in Solana and Algorand smart contracts?}
\end{mdframed}

The objective of our work is to address all the outlined RQs by first introducing the current vulnerabilities and analysis tools for Algorand and Solana, then by evaluating the effectiveness of prompt engineering techniques—specifically zero-shot and few-shot—and finally by applying fine-tuning strategies to further enhance LLM-based vulnerability assessments. This multi-layered approach enables a broader and more precise understanding of how language models can be leveraged in emerging blockchain ecosystems.

\section{Experimental Evaluation}

In this section, we present the response to the outlined RQs by first introducing the most critical vulnerabilities currently existing in Algorand and Solana blockchains, answering RQ1. Subsequently, in sections \ref{resp_rq2}, and \ref{resp_rq3} we analyze the role of LLM in detecting those vulnerabilities, and the design of fine-tuning and prompt engineering, respectively.
\begin{table}[h!]
\centering
\caption{Smart Contract Vulnerabilities according to OWASP Top 10 (2025) \cite{owasp_top10_2025}}
\begin{tabular}{|c|l|c|}
\hline
\textbf{ID} & \textbf{Vulnerability} \\
\hline
$V_1$  & Access Control Vulnerabilities \\
$V_2$  & Price Oracle Manipulation \\
$V_3$  & Logic Errors\\
$V_4$  & Lack of Input Validation\\
$V_5$  & Reentrancy Attacks \\
$V_6$  & Unchecked External Calls\\
$V_7$  & Flash Loan Attacks \\
$V_8$  & Integer Overflow and Underflow\\
$V_9$  & Insecure Randomness\\
$V_{10}$ & Denial of Service (DoS) Attacks\\
\hline
\end{tabular}
\label{tab:sc_vulnerabilities}
\end{table}

\subsection{RQ1 – Vulnerabilities in Algorand and Solana}
To analyze the vulnerabilities in Algorand and Solana blockchains, we take as a reference of the relevance of vulnerabilities the recent OWASP Top 10, depicted in Fig. \ref{tab:sc_vulnerabilities}. By taking this as a reference, we try to give a pseudo-mapping of these vulnerabilities over the existing vulnerabilities of Algorand and Solana blockchains.
 \subsubsection{Algorand}
Algorand leverages the Transaction Execution Approval Language (TEAL) for the creation and execution of smart contracts. TEAL is designed to prioritize both scalability and security while enhancing transaction performance. Despite enhancements to blockchain structures, smart contracts are susceptible to most of the vulnerabilities found in Ethereum smart contracts.

$V_1$ maps to \textit{Arbitrary Update} and \textit{Arbitrary Delete}, as both involve state-modifying calls that can be dangerous if executed without explicit verification of the caller’s identity. Additionally, \textit{Unchecked Payment Receiver} and \textit{Unchecked Asset Receiver} fall under this category, since the lack of validation on receiving entities can lead to unauthorized fund or asset transfers. $V_2$ is feasible in Algorand when smart contracts rely on external data sources without proper authentication or data integrity guarantees. As Algorand lacks a built-in oracle mechanism, developers typically use off-chain services to input external data (e.g., price feeds). Thus, the vulnerability arises not from the protocol itself, but from insecure oracle integration. For this reason, no direct mapping can be applied within the Algorand context. $V_3$ pertains to logical vulnerabilities—issues that stem from how the smart contract logic is written rather than specific instructions, hence not directly covered by this study. $V_5$ is not applicable in Algorand. All transactions and transaction groups are atomic and stateless beyond their execution scope. Since Algorand does not support dynamic callbacks or inter-contract call stacks like Ethereum, this class of vulnerability is effectively prevented. $V_6$ appears in Algorand as \textit{Unchecked RekeyTo}, which allows dangerous account-level permission changes if the \texttt{rekey\_to} field is not explicitly restricted or validated. $V_7$ is theoretically possible but practically difficult to exploit in Algorand due to the platform’s architectural features. Atomic transaction groups, absence of transaction nonces, and lack of penalties for failed transactions make flash loan-based attacks far less feasible. As a result, Algorand offers a safer environment for using flash loans \cite{algorand_flashloans}. $V_8$ relates to TEAL-level programming logic. Vulnerabilities such as \textit{Arithmetic Overflow/Underflow} and \textit{Unchecked Transaction Fee} are typical examples, reflecting low-level arithmetic errors or unsafe assumptions about numerical operations. $V_9$ reflects vulnerabilities; an insecure source is used as randomness. In Algorand, randomness can be guaranteed through the usage of Verifiable Random Function (VRF) \cite{esgin2021practical}, hence preventing this vulnerability. $V_{10}$ is only partially mitigated in Algorand. The protocol architecture limits information disclosure to the block producer, reducing the attack surface for targeted DoS attacks. However, application-level DoS is still possible, particularly through \textit{Unchecked Transaction Fee} abuse, where users can exhaust system resources by shifting fee burdens onto others.
\begin{table}[h!]
\centering
\caption{Mapping of Vulnerabilities ($V_1$–$V_{10}$) in Algorand and Solana}
\begin{tabular}{|p{0.3cm}|p{3.5cm}|p{3.5cm}|}
\hline
\textbf{ID} & \textbf{Algorand} & \textbf{Solana} \\
\hline
$V_1$ & \textit{Arbitrary Update, Arbitrary Delete}, \textit{Unchecked Payment Receiver}, \textit{Unchecked Asset Receiver} & Access control vulnerabilities are applicable. Requires \textit{Owner Check}, \textit{Signer Check}, \textit{Key Check}. \\
\hline
$V_2$ & Vulnerability arises from off-chain oracle integration without authentication. Out of scope. & Oracle manipulation is possible via oracles like Pyth or Switchboard. Out of scope. \\
\hline
$V_3$ & Generic logical vulnerabilities. Out of scope. & Generic logical vulnerabilities. Out of scope. \\
\hline
$V_4$ & N/A. & \textit{Type Confusion} due to missing type checks when parsing account inputs. \\
\hline
$V_5$ & N/A. Algorand's atomic, stateless transaction model prevents this class & \textit{Cross-Program Invocation} (CPI) introduces risks similar to reentrancy. Requires strict validation. \\
\hline
$V_6$ & Present as \textit{Unchecked RekeyTo} (dangerous rekeying without validation) & Tied to \textit{Bump Seed}; unchecked low-level calls possible if not validated. \\
\hline
$V_7$ & N/A. Hard to exploit due to atomic groups, no nonce, and no penalties for failed txs. & Flash loan attacks are possible but depend on external protocols. Out of scope. \\
\hline
$V_8$ & TEAL-specific: Arithmetic overflow/underflow, Unchecked Transaction Fee & Integer overflow/underflow risk if no validation is performed. \\
\hline
$V_9$ & N/A. & PDA collisions from \textit{Bump Seed}; similar to $V_6$. \\
\hline
$V_{10}$ & Partially mitigated; DoS possible via \textit{Unchecked Transaction Fee} and resource exhaustion & Leader election DoS possible if randomness is predictable; network disruption could occur. \\
\hline
\end{tabular}
\label{tab:algorand_solana_vulnerabilities}
\end{table}

\subsubsection{Solana}
A notable distinction between Solana and other blockchain platforms lies in its strong resistance to time manipulation and front-running attacks. Solana’s Proof of History (PoH) consensus mechanism generates a cryptographically verifiable timeline of events, making it difficult for malicious actors to tamper with timestamps or influence transaction order.

$V_1$ is relevant to the Solana blockchain, as well. Effective access control mechanisms—such as verifying ownership, signer status, and key consistency through techniques like \textit{Owner Check}, \textit{Signer Check}, and \textit{Key Check}—are essential to prevent unauthorized access to program state and accounts. $V_2$ and $V_7$ are common threats in the Solana ecosystem, as discussed in \cite{wu2025exploring}. However, both oracle manipulation ($V_2$) and flash loan-based exploits ($V_7$) primarily arise from external dependencies, such as data feeds or liquidity protocols, rather than from the smart contract code itself. As such, these vulnerabilities fall outside the primary scope of this study. $V_3$ concerns logical vulnerabilities—flaws in how a program is structured or designed, independent of specific low-level instructions. These are implementation-level issues that, while critical, are not directly addressed in this analysis. $V_4$ corresponds to account confusion, a risk that occurs when a smart contract accepts input accounts without proper validation. In Solana, this vulnerability arises when a program incorrectly interprets account data. Programs typically expect a specific data layout and may fail to verify that the input account matches the required type or ownership, resulting in unexpected behavior or security breaches. Solana’s \textit{Cross-Program Invocation} (CPI) introduces unique risks analogous to vulnerabilities like reentrancy in platforms such as Ethereum. These risks map to $V_5$. CPI allows a program to invoke other programs, but it lacks built-in validation. If the calling program fails to enforce strict controls over these invocations, attackers may exploit the interaction path. Similarly, \textit{Bump Seed}, which is close to $V_6$, represents a bad management of Program Delivery Address (PDA), which can lead to executing external calls without a check. $V_8$ covers issues such as integer overflow and underflow, which remain relevant on Solana, similar to their presence in Ethereum. These arithmetic vulnerabilities can be used to subvert program logic or bypass limits if appropriate validation checks are omitted. $V_9$ reflects risks tied to blockchain infrastructure rather than smart contract logic. For instance, Program Derived Addresses (PDAs) in Solana—used to create contract-specific accounts—could be susceptible to collision or forgery if constructed using weak or predictable seeds. This is comparable to a “short address attack,” in which an attacker could derive an identical PDA, enabling them to hijack associated resources or transactions. It is possible to refer to it as \textit{Bump Seed}. $V_{10}$ could potentially be exploited through Solana’s leader election process. If an attacker is able to predict or influence the randomness used in validator selection, they might anticipate upcoming block producers and launch a targeted Denial-of-Service (DoS) attack against them. This could impair block production and affect transaction processing across the network.

\subsubsection{Discussion}
A detailed mapping of the most critical vulnerabilities—originally identified in Ethereum—has been presented in Table~\ref{tab:algorand_solana_vulnerabilities}, with a comparative analysis for Algorand and Solana. Among them, access control vulnerabilities (corresponding to $V_1$), which top the OWASP ranking, remain highly relevant across both platforms. In both Algorand and Solana, insufficient validation of critical transaction fields (such as \texttt{sender}, \texttt{rekey\_to}, or asset transfer targets) leads to similar exploit opportunities. However, the inherent architectural features of Algorand prevent four out of the ten vulnerabilities analyzed from being applicable. This highlights the robustness of Algorand’s design in mitigating certain classes of attacks by construction \cite{gilad2017algorand}. In contrast, all ten vulnerabilities remain theoretically applicable in Solana, though their practical exploitability is often constrained by the platform’s complexity and parallel execution model. For example, in the case of $V_{10}$ (denial of service via leader targeting), the exploit would require a highly sophisticated orchestration capable of predicting the randomness governing leader election—a significantly more difficult task than in Ethereum. Notably, only $V_1$ (Access Control) and $V_6$ (Unchecked External Calls) present a clear one-to-one mapping across all three platforms, suggesting that a uniform security model cannot be applied to non-EVM blockchains, but a baseline of those can be generalized for all the decentralized environments. The analysis confirms the need for platform-specific threat models and security analyses tailored to the execution semantics and consensus architecture of each blockchain environment.

\subsection{RQ2 – LLMs and Domain-Specific Languages}
\label{resp_rq2}

To address RQ2, we first describe the construction of the dataset and then present the main findings derived from evaluating two LLMs across both blockchains. For each blockchain, we selected a representative set of vulnerabilities as defined in RQ1, and designed a uniform prompt for querying the LLMs. Responses were collected and labeled according to a binary classification scheme according to the specific category of that vulnerability. The datasets were built manually based on the OWASP-inspired vulnerability taxonomy, supported by symbolic and structural analysis. For each vulnerability type, at least five vulnerable and five non-vulnerable instances were developed, resulting in balanced datasets for both PyTeal (Algorand) and Rust (Solana). These examples were crafted to reflect realistic contract behaviors while isolating the vulnerability patterns under investigation. We selected two open-source and tunable LLMs for evaluation: LLAMA-3-8B and DeepSeek-R1-Distill-Qwen-14B. These models were chosen due to their strong open-access communities and compatibility with both inference and fine-tuning, which is not always feasible with proprietary advanced models (\emph{e.g. GPT-4}).

\subsubsection{Experimental Setup}

To isolate the model's baseline detection ability—excluding any prompt engineering techniques—we used a simple and consistent query format across all examples:

\begin{tcolorbox}[colback=gray!5, colframe=gray!80, title=System Prompt]
\texttt{<SYS>} Can you check if the following smart contract written in [Programming Language] contains a vulnerability? --[Source Code]--\texttt{</SYS>}
\end{tcolorbox}

This prompt has been inserted each time on a freshly loaded model, so that each response is not influenced by the previous ones. We assessed model responses using standard evaluation metrics: accuracy, precision, recall, and F1-score. To mitigate the effect of stochasticity in LLM outputs, we executed each prompt three times per vulnerability type and averaged the results. 
To assess the accuracy of the model, we leveraged a binary approach: if the model takes as input a smart contract vulnerable to $V_x$ and classifies it as $V_y$ this will be taken as a false negative; hence, only correctly assessed vulnerabilities will be assigned to TP. The overall testing phase consists of 3 requests for each positive and negative sample, resulting in 30 requests for each vulnerability, demonstrating its reliability against randomness introduced by LLM.

\begin{table*}[ht]
\centering
\caption{Comparison of vulnerability detection performance for Solana and Algorand between DeepSeek (DS) and LLaMA (LM).}
\begin{tabular}{|c|l|cc|cc|cc|cc|}
\hline
\multirow{2}{*}{\textbf{Blockchain}} & \multirow{2}{*}{\textbf{Vulnerability}} 
& \multicolumn{2}{c|}{\textbf{Accuracy}} 
& \multicolumn{2}{c|}{\textbf{Precision}} 
& \multicolumn{2}{c|}{\textbf{F1-score}} 
& \multicolumn{2}{c|}{\textbf{Recall}} \\
\cline{3-10}
& & DS & LM & DS & LM & DS & LM & DS & LM \\
\hline

\multirow{5}{*}{Solana} 
& Bump Seed         & 0.80 & 0.60 & 1.00 & 1.00 & 0.75 & 0.33 & 0.60 & 0.20 \\
& CPI               & 0.53 & 0.60 & 0.60 & 1.00 & 0.30 & 0.33 & 0.20 & 0.20 \\
& Integer Flow      & 0.67 & 0.43 & 0.78 & 0.00 & 0.58 & 0.00 & 0.47 & 0.00 \\
& Missing Key Check & 0.57 & 0.50 & 1.00 & 0.00 & 0.24 & 0.00 & 0.13 & 0.00 \\
& Type Confusion    & 0.60 & 0.50 & 1.00 & 0.00 & 0.33 & 0.00 & 0.20 & 0.00 \\
\hline
& \textbf{Avg.}     & \textbf{0.63} & \textbf{0.53} &  &  &  &  &  & \\
\hline
\hline

\multirow{8}{*}{Algorand} 
& Arbitrary Deletion           & 0.57 & 0.50 & 1.00 & 0.00 & 0.24 & 0.00 & 0.13 & 0.00 \\
& Arbitrary Update             & 0.73 & 0.50 & 1.00 & 0.00 & 0.64 & 0.00 & 0.47 & 0.00 \\
& Unchecked Asset Close To     & 0.50 & 0.50 & 0.00 & 0.00 & 0.00 & 0.00 & 0.00 & 0.00 \\
& Unchecked Close Remainder To & 0.53 & 0.50 & 1.00 & 0.00 & 0.13 & 0.00 & 0.07 & 0.00 \\
& Unchecked Rekey To           & 0.57 & 0.50 & 1.00 & 0.00 & 0.24 & 0.00 & 0.13 & 0.00 \\
& Unchecked Transaction Fee    & 0.57 & 0.50 & 1.00 & 0.00 & 0.24 & 0.00 & 0.13 & 0.00 \\
& Unchecked Asset Receiver     & 0.60 & 0.50 & 1.00 & 0.00 & 0.33 & 0.00 & 0.20 & 0.00 \\
& Unchecked Payment Receiver   & 0.70 & 0.50 & 1.00 & 0.00 & 0.57 & 0.00 & 0.40 & 0.00 \\
\hline
& \textbf{Avg.}     & \textbf{0.60} & \textbf{0.50} &  &  &  &  &  & \\
\hline
\end{tabular}
\label{tab:no_tuning}
\end{table*}

\subsubsection{Discussion}

The results shown in Table~\ref{tab:no_tuning} reveal several important trends in LLM-based vulnerability detection across the Solana and Algorand smart contract datasets. Overall, DeepSeek consistently outperforms LLaMA in terms of detection performance across all evaluation metrics. 

In the case of Solana, DeepSeek achieves an average accuracy of 0.63 compared to LLaMA's 0.53. Notably, DeepSeek reaches perfect precision in all five tested vulnerabilities, indicating that when it identifies a vulnerability, it does so with confidence. However, its recall varies substantially—ranging from 0.20 (Type Confusion) to 0.60 (Bump Seed)—suggesting inconsistency in detecting all vulnerable instances. LLaMA, while more conservative, fails to detect most true positives, often producing zero recall and F1-score across several vulnerabilities. For Algorand, DeepSeek again outperforms LLaMA across the board, with an average accuracy of 0.60 compared to LLaMA’s 0.50. The disparity in F1-score is especially stark: DeepSeek shows moderate detection capability (e.g., F1 = 0.64 for Arbitrary Update, F1 = 0.57 for Unchecked Payment Receiver), while LLaMA yields a null F1-score in all categories, due to its complete lack of recall. This highlights the model’s inability to capture vulnerability patterns in PyTeal-based smart contracts when used without fine-tuning. The performance gap between platforms also reflects the underlying complexity of their domain-specific languages. While Solana’s Rust-based contracts seem to be more detectable by both models, Algorand’s PyTeal contracts appear more challenging due to their stack-based logic and less expressive syntax. This emphasizes the need for platform-specific fine-tuning or prompt calibration to achieve robust results in non-EVM contexts. These findings support the hypothesis that LLMs can serve as lightweight vulnerability detection tools, but their performance is highly sensitive to the language and structure of the contract, as well as the model configuration. Without prompt engineering or training adaptation, models like LLaMA fail to generalize, whereas DeepSeek shows moderate out-of-the-box competence. In the next section, we will introduce our approach to fine-tuning and prompt engineering as a strategy to improve the metrics.

\subsection{RQ3 – Prompt Engineering vs Fine-Tuning}
\label{resp_rq3}

\begin{table}[h!]
\centering
\caption{Comparison of Vulnerability Detection Accuracy for Solana and Algorand Across Fine-Tuning (FT), Prompt Engineering (PE), and their Combination Between DeepSeek (DS) and LLaMA (LM).}
\begin{tabular}{|p{2.3cm}|c|c|c|c|c|c|}
\hline\multirow{2}{*}{\textbf{Vulnerability}} & \multicolumn{2}{c|}{\textbf{Fine-Tuning}} & \multicolumn{2}{c|}{\textbf{Prompt Eng.}} & \multicolumn{2}{c|}{\textbf{PE + FT}} \\
\cline{2-7}
 & DS & LM & DS & LM & DS & LM \\
\hline
\rowcolor{gray!20}
\multicolumn{7}{|c|}{Solana}\\
\hline
Bump Seed & 0.50 & 0.56 & 0.63 & 0.60 & 0.50 & 0.60 \\
CPI & 0.67 & 0.73 & 0.63 & 0.53 & 0.63 & 0.70 \\
Integer Flow & 0.43 & 0.53 & 0.53 & 0.50 & 0.53 & 0.43 \\
Missing Key Check & 0.76 & 0.67 & 0.63 & 0.67 & 0.73 & 0.77 \\
Type Confusion & 0.50 & 0.63 & 0.73 & 0.50 & 0.63 & 0.53 \\
\hline
\textbf{Avg.} & \textbf{0.57} & \textbf{0.62} & \textbf{0.63} & \textbf{0.56} & \textbf{0.60} & \textbf{0.61} \\
\hline
\hline
\rowcolor{gray!20}
\multicolumn{7}{|c|}{Algorand}\\
\hline
 
Arbitrary Del. & 0.57 & 0.50 & 0.60 & 0.50 & 0.60 & 0.67 \\
Arbitrary Upd. & 0.67 & 0.50 & 0.60 & 0.50 & 0.67 & 0.57 \\
Unc. Asset Close To & 0.60 & 0.50 & 0.50 & 0.50 & 0.43 & 0.47 \\
Unc. Close Remainder & 0.60 & 0.50 & 0.50 & 0.50 & 0.53 & 0.60 \\
Unc. Rekey To & 0.57 & 0.50 & 0.53 & 0.50 & 0.43 & 0.67 \\
Unc. TX Fee & 0.63 & 0.50 & 0.50 & 0.50 & 0.53 & 0.83 \\
Unc. Asset Rec. & 0.73 & 0.50 & 0.53 & 0.50 & 0.70 & 0.83 \\
Unc. Payment Rec. & 0.50 & 0.53 & 0.50 & 0.50 & 0.70 & 0.60 \\
\hline
\textbf{Avg.} & \textbf{0.62} & \textbf{0.50} & \textbf{0.54} & \textbf{0.50} & \textbf{0.57} & \underline{\textbf{0.65}} \\
\hline
\end{tabular}
\label{tab:tuning}
\end{table}
\normalsize
\subsubsection{Prompt Design}
We devised a clear prompt belonging to the role-based prompting.

\begin{tcolorbox}[colback=gray!5, colframe=gray!80, title=System Prompt]
\texttt{<SYS>} You are a smart contract security analyzer. You receive smart contract
written in [Programming Language] as input and answer with the vulnerability
identified if exist. The vulnerabilities are classified according to OWASP Top 10.\texttt{</SYS>}
\end{tcolorbox}
Then we re-used the same system prompt explained in the experimental setup.

\subsubsection{Fine-Tuning}
For the fine-tuning of the pre-trained model, we used the same approach for both LLaMA and DeepSeek models. We fine-tuned the models using the Hugging Face \texttt{transformers} and \texttt{peft} libraries in conjunction with LoRA (Low-Rank Adaptation) to reduce the number of trainable parameters and memory footprint. This approach enables efficient fine-tuning on consumer-grade hardware while preserving base model generalization. DeepSeek has been trained for 10 epochs, while LLaMA has been trained for 4 epochs. We used the same LoRA parameters for both models: 64 for $r$, 16 $\alpha$, and 0.1 dropout. We used an Adam optimizer with learning rate = 2e-4 and maximum sequence length of 2048 tokens. To standardize output evaluation, we applied a post-processing rule that maps the generated token to the binary class label.
\subsubsection{Discussion}

The results summarized in Table~\ref{tab:tuning} reveal several trends regarding the detection capabilities of LLMs across blockchain platforms and training strategies. First, DeepSeek consistently outperforms LLaMA across most configurations. In Solana, DeepSeek achieves the highest accuracy with prompt engineering alone (0.63), while in Algorand, its best result comes from fine-tuning (0.62). On the other hand, LLaMA exhibits its strongest performance when prompt engineering is combined with fine-tuning, particularly in Algorand, where it reaches an accuracy of 0.65—its peak across all configurations. These findings suggest that while DeepSeek benefits from architectural robustness and generalizability, LLaMA is more sensitive to training context and requires more extensive guidance to achieve competitive performance.

Additionally, prompt engineering alone proves highly effective, especially for DeepSeek, indicating that structured prompts can yield strong baselines without additional model adaptation. This is particularly evident in Solana, where prompt-based querying achieves near-equivalent results to fine-tuning, despite the absence of additional training.

\begin{figure}
    \centering
    \includegraphics[width=0.9\linewidth]{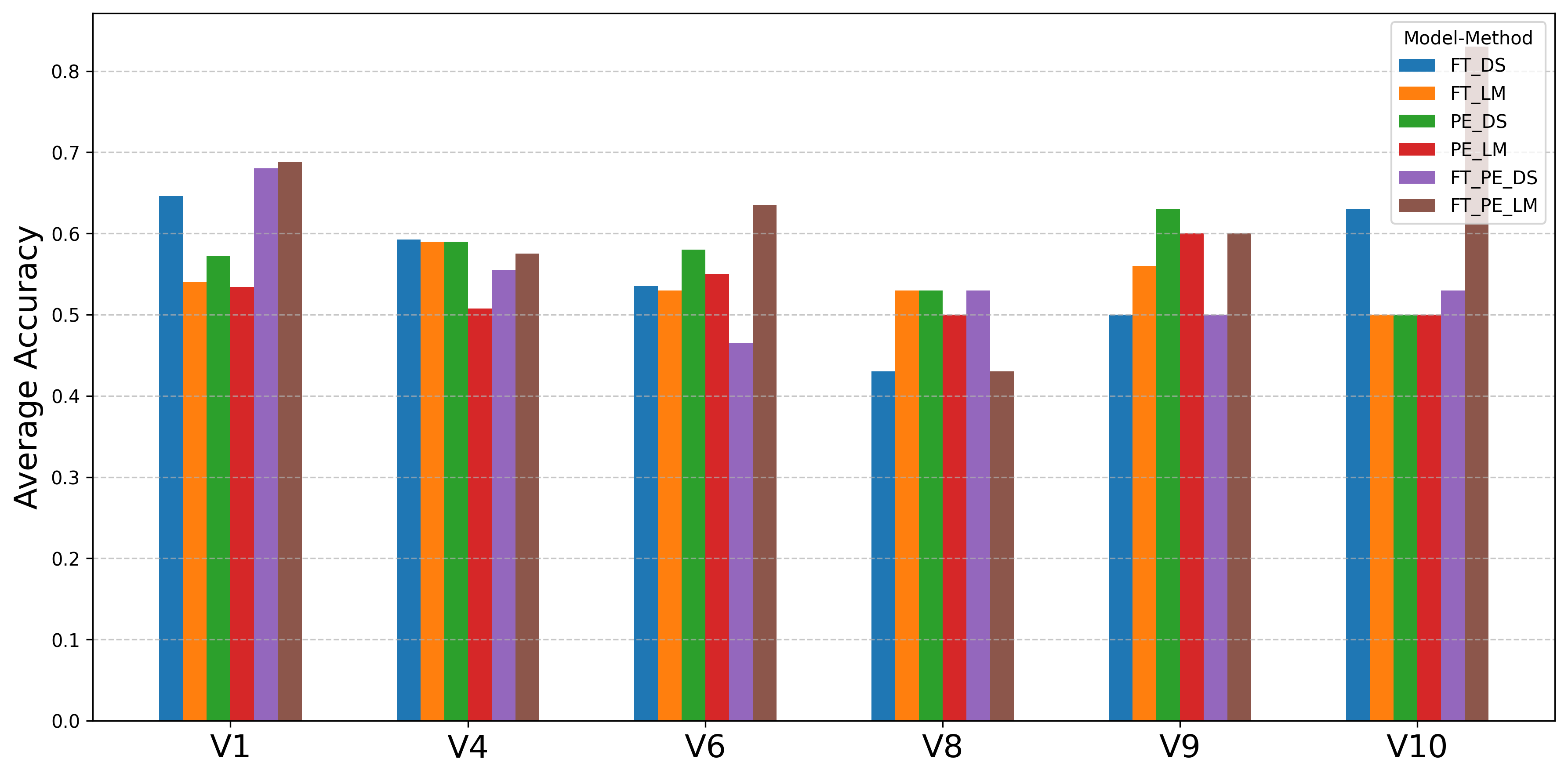}
    \caption{Comparison of Vulnerability Type Accuracy for Solana and Algorand Across Fine-Tuning (FT), Prompt Engineering (PE), and their Combination (FT\_PE) Between DeepSeek (DS) and LLaMA (LM).}
    \label{fig:group_by_vuln}
\end{figure}

Certain vulnerabilities appear easier to detect, such as \textit{MissingKeyCheck} in Solana and \textit{UncheckedAssetReceiver} in Algorand, which exhibit high accuracy across all configurations. Conversely, more subtle or logic-dependent patterns (e.g., \textit{UncheckedCloseRemainder}, \textit{TypeConfusion}) show greater variability, indicating that model performance is tightly linked to how explicitly the vulnerability is encoded in the source. This behavior is confirmed by Fig. \ref{fig:group_by_vuln}, where $V_1$ and $V_{10}$ are the easiest to detect vulnerabilities. Algorand’s PyTeal contracts, being stack-based and less expressive than Solana’s Rust contracts, seem to benefit more from hybrid training strategies. This is reflected in the relatively higher gains for LLaMA under the Prompt and Fine-Tuning setting. In summary, the analysis suggests the following:
\begin{itemize}
  \item DeepSeek offers stable and robust performance across settings, particularly, independently from the prompt used;
  \item LLaMA improves significantly with the addition of prompt engineering to fine-tuning,
  \item Prompt engineering alone remains a strong baseline for vulnerability detection, reducing reliance on costly fine-tuning.
\end{itemize}

\section{Conclusion}

In this work, we explored the applicability of LLMs for detecting vulnerabilities in smart contracts beyond the Ethereum Virtual Machine (EVM) paradigm. Our experimental results show that prompt engineering provides a competitive and cost-effective baseline, particularly in semantically rich languages like Rust (Solana). Fine-tuning offers benefits primarily when dealing with abstract, lower-level languages such as TEAL (Algorand), where model understanding is limited by structural constraints. The hybrid configuration proved most effective in such cases, particularly for LLaMA, which achieved its best accuracy when prompt engineering was combined with fine-tuning. However, effectiveness is closely tied to both the linguistic expressiveness of the target DSL and the adaptation strategy employed. Future work will explore the integration of addiction techniques of prompt-engineering (\emph{e.g. few-shot}) and larger datasets for more precise evaluation of the models.

\section*{Acknowledgment}
This work was partially supported by project SERICS (PE00000014) under the NRRP MUR program funded by the EU - NGEU.

\bibliographystyle{IEEEtran}
\bibliography{refs}

@article{cai2023impact,
  title={The impact of perceived benefits on blockchain adoption in supply chain management},
  author={Cai, Chuangneng and Hao, Xiancheng and Wang, Kui and Dong, Xuebing},
  journal={Sustainability},
  volume={15},
  number={8},
  pages={6634},
  year={2023},
  publisher={MDPI}
}

@inproceedings{hammoudeh2023digital,
  title={Digital Certificate Validation Using Blockchain: A Survey},
  author={Hammoudeh, Yazan Abu and Qatawneh, Mohammad and AbuAlghanam, Orieb and Almaiah, Mohammed A},
  booktitle={2023 International Conference on Information Technology (ICIT)},
  pages={506--510},
  year={2023},
  organization={IEEE}
}

@inproceedings{napoli2024leveraging,
  title={Leveraging Large Language Models for Automatic Smart Contract Generation},
  author={Napoli, Emanuele Antonio and Barb{\`a}ra, Fadi and Gatteschi, Valentina and Schifanella, Claudio},
  booktitle={2024 IEEE 48th Annual Computers, Software, and Applications Conference (COMPSAC)},
  pages={701--710},
  year={2024},
  organization={IEEE}
}

@article{bartoletti2024smart,
  title={Smart Contract Languages: a comparative analysis},
  author={Bartoletti, Massimo and Benetollo, Lorenzo and Bugliesi, Michele and Crafa, Silvia and Sasso, Giacomo Dal and Pettinau, Roberto and Pinna, Andrea and Piras, Mattia and Rossi, Sabina and Salis, Stefano and others},
  journal={arXiv preprint arXiv:2404.04129},
  year={2024}
}

@inproceedings{sun2023panda,
  title={Panda: Security analysis of algorand smart contracts},
  author={Sun, Zhiyuan and Luo, Xiapu and Zhang, Yinqian},
  booktitle={32nd USENIX Security Symposium (USENIX Security 23)},
  pages={1811--1828},
  year={2023}
}

@article{poston2020mapping,
  title={Mapping the OWASP top ten to blockchain},
  author={Poston, Howard},
  journal={Procedia Computer Science},
  volume={177},
  pages={613--617},
  year={2020},
  publisher={Elsevier}
}

@incollection{bhardwaj2021attack,
  title={Attack vectors for blockchain and mapping OWASP vulnerabilities to smart contracts},
  author={Bhardwaj, Akashdeep and Goundar, Sam},
  booktitle={Blockchain Technologies, Applications and Cryptocurrencies: Current Practice and Future Trends},
  pages={139--156},
  year={2021},
  publisher={World Scientific}
}

@misc{owasp_top10_2025,
  author       = {{OWASP Foundation}},
  title        = {{OWASP Smart Contract Top 10 (2025)}},
  year         = {2025},
  howpublished = {\url{https://owasp.org/www-project-smart-contract-top-10/}},
  note         = {Accessed: 2025-03-11}
}

@inproceedings{estrada2023systematic,
  title={A Systematic Literature Review of Blockchain Technology: Applications Fields, Platforms, and Consensus Protocols},
  author={Estrada, Carlos A and Naranjo, Sergio S and Toasa, Veronica J and Yoo, Sang Guun},
  booktitle={Proceedings of the 2023 7th International Conference on Computer Science and Artificial Intelligence},
  pages={123--131},
  year={2023}
}

@inproceedings{karanjai2024solmover,
  title={SolMover: Feasibility of Using LLMs for Translating Smart Contracts},
  author={Karanjai, Rabimba and Xudagger, Lei and Shi, Weidong},
  booktitle={2024 IEEE International Conference on Blockchain and Cryptocurrency (ICBC)},
  pages={1--3},
  year={2024},
  organization={IEEE}
}

@INPROCEEDINGS{10456393,
  author={Akuthota, Vishwanath and Kasula, Raghunandan and Sumona, Sabiha Tasnim and Mohiuddin, Masud and Reza, Md Tanzim and Rahman, Md Mizanur},
  booktitle={2023 IEEE 9th International Women in Engineering (WIE) Conference on Electrical and Computer Engineering (WIECON-ECE)}, 
  title={Vulnerability Detection and Monitoring Using LLM}, 
  year={2023},
  volume={},
  number={},
  pages={309-314},
  doi={10.1109/WIECON-ECE60392.2023.10456393}}

@article{rautenberg2022case,
  title={A Case Study of Security Vulnerabilities in Smart Contracts},
  author={Rautenberg, Marvin James and Rezabek, Filip},
  journal={Network},
  volume={53},
  year={2022}
}

@inproceedings{de2022evaluating,
  title={Evaluating Blockchain Systems: A Comprehensive Study of Security and Dependability Attributes.},
  author={De Angelis, Stefano and Zanfino, Gilberto and Aniello, Leonardo and Lombardi, Federico and Sassone, Vladimiro},
  booktitle={DLT@ ITASEC},
  pages={18--32},
  year={2022}
}

@article{deangelis2023,
  title={Security and dependability analysis of blockchain systems in partially synchronous networks with Byzantine faults},
  author={De Angelis, Stefano and Lombardi, Federico and Zanfino, Gilberto and Aniello, Leonardo and Sassone, Vladimiro},
  journal={International Journal of Parallel, Emergent and Distributed Systems},
  pages={1--21},
  year={2023},
  publisher={Taylor \& Francis}
}

@inproceedings{mothukuri2024llmsmartsec,
  title={LLMSmartSec: Smart Contract Security Auditing with LLM and Annotated Control Flow Graph},
  author={Mothukuri, Viraaji and Parizi, Reza M and Massa, James L},
  booktitle={2024 IEEE International Conference on Blockchain (Blockchain)},
  pages={434--441},
  year={2024},
  organization={IEEE}
}

@INPROCEEDINGS{10664408,
  author={Luo, Yu and Xu, Weifeng and Andersson, Karl and Hossain, Mohammad Shahadat and Xu, Dianxiang},
  booktitle={2024 IEEE International Conference on Blockchain (Blockchain)}, 
  title={FELLMVP: An Ensemble LLM Framework for Classifying Smart Contract Vulnerabilities}, 
  year={2024},
  volume={},
  number={},
  pages={89-96},
  doi={10.1109/Blockchain62396.2024.00021}}

@article{boi2024smart,
  title={Smart Contract Vulnerability Detection: The Role of Large Language Model (LLM)},
  author={Boi, Biagio and Esposito, Christian and Lee, Sokjoon},
  journal={ACM SIGAPP Applied Computing Review},
  volume={24},
  number={2},
  pages={19--29},
  year={2024},
  publisher={ACM New York, NY, USA}
}

@article{soud2024soley,
  title={Soley: Identification and automated detection of logic vulnerabilities in ethereum smart contracts using large language models},
  author={Soud, Majd and Nuutinen, Waltteri and Liebel, Grischa},
  journal={arXiv preprint arXiv:2406.16244},
  year={2024}
}

@article{radcliffe2024automated,
  title={Automated prompt engineering for semantic vulnerabilities in large language models},
  author={Radcliffe, Thomas and Lockhart, Emily and Wetherington, James},
  journal={Authorea Preprints},
  year={2024}
}

@INPROCEEDINGS{10698605,
  author={Shenoy, Neethu and Mbaziira, Alex V},
  booktitle={2024 International Conference on Electrical, Computer and Energy Technologies (ICECET}, 
  title={An Extended Review: LLM Prompt Engineering in Cyber Defense}, 
  year={2024},
  volume={},
  number={},
  pages={1-6},
  doi={10.1109/ICECET61485.2024.10698605}}

@Article{electronics13132657,
AUTHOR = {Bae, Jaehyeon and Kwon, Seoryeong and Myeong, Seunghwan},
TITLE = {Enhancing Software Code Vulnerability Detection Using GPT-4o and Claude-3.5 Sonnet: A Study on Prompt Engineering Techniques},
JOURNAL = {Electronics},
VOLUME = {13},
YEAR = {2024},
NUMBER = {13},
ARTICLE-NUMBER = {2657},
URL = {https://www.mdpi.com/2079-9292/13/13/2657},
ISSN = {2079-9292},
DOI = {10.3390/electronics13132657}
}

@INPROCEEDINGS{10479384,
  author={Lu, Guilong and Ju, Xiaolin and Chen, Xiang and Yang, Shaoyu and Chen, Liang and Shen, Hao},
  booktitle={2023 30th Asia-Pacific Software Engineering Conference (APSEC)}, 
  title={Assessing the Effectiveness of Vulnerability Detection via Prompt Tuning: An Empirical Study}, 
  year={2023},
  volume={},
  number={},
  pages={415-424},
  doi={10.1109/APSEC60848.2023.00052}}

@article{ortu4530467identifying,
  title={Identifying and fixing vulnerable patterns in ethereum smart contracts: A comparative study of fine-tuning and prompt engineering using large language models},
  author={Ortu, Marco and Ibba, Giacomo and Conversano, Claudio and Tonelli, Roberto and Destefanis, Giuseppe},
  journal={Available at SSRN 4530467}
}

@INPROCEEDINGS{10679877,
  author={Ma, Jiarun and Feng, Shiling and Zeng, Jiahao and Lu, Jia and Chen, Jie},
  booktitle={2024 International Conference on Networking and Network Applications (NaNA)}, 
  title={Smart Contract Vulnerability Detection Based on Prompt-guided ChatGPT}, 
  year={2024},
  volume={},
  number={},
  pages={321-326},
  doi={10.1109/NaNA63151.2024.00060}}

@inproceedings{tikhomirov2018smartcheck,
  title={Smartcheck: Static analysis of ethereum smart contracts},
  author={Tikhomirov, Sergei and Voskresenskaya, Ekaterina and Ivanitskiy, Ivan and Takhaviev, Ramil and Marchenko, Evgeny and Alexandrov, Yaroslav},
  booktitle={Proceedings of the 1st international workshop on emerging trends in software engineering for blockchain},
  pages={9--16},
  year={2018}
}

@inproceedings{feist2019slither,
  title={Slither: a static analysis framework for smart contracts},
  author={Feist, Josselin and Grieco, Gustavo and Groce, Alex},
  booktitle={2019 IEEE/ACM 2nd International Workshop on Emerging Trends in Software Engineering for Blockchain (WETSEB)},
  pages={8--15},
  year={2019},
  organization={IEEE}
}

@INPROCEEDINGS{10740107,
  author={Fu, Yihang and Jing, Mingwei and Zhou, Jiaolun and Wu, Peilin and Wang, Ye and Zhang, Luyao and Hu, Chuang},
  booktitle={2024 IEEE International Conference on Metaverse Computing, Networking, and Applications (MetaCom)}, 
  title={Quantifying the Blockchain Trilemma: A Comparative Analysis of Algorand, Ethereum 2.0, and Beyond}, 
  year={2024},
  volume={},
  number={},
  pages={97-104},
  doi={10.1109/MetaCom62920.2024.00028}}

@misc{algorand_flashloans,
  author       = {{Algorand Developer}},
  title        = {{AVM, EVM \& Flash Loans}},
  howpublished = {\url{https://developer.algorand.org/solutions/avm-evm-flash-loans/}},
  note         = {Accessed: 2025-03-13},
  year         = {n.d.}
}

@article{wu2025exploring,
  title={Exploring Vulnerabilities and Concerns in Solana Smart Contracts},
  author={Wu, Xiangfan and Xing, Ju and Li, Xiaoqi},
  journal={arXiv preprint arXiv:2504.07419},
  year={2025}
}

@inproceedings{esgin2021practical,
  title={Practical post-quantum few-time verifiable random function with applications to algorand},
  author={Esgin, Muhammed F and Kuchta, Veronika and Sakzad, Amin and Steinfeld, Ron and Zhang, Zhenfei and Sun, Shifeng and Chu, Shumo},
  booktitle={International Conference on Financial Cryptography and Data Security},
  pages={560--578},
  year={2021},
  organization={Springer}
}

@inproceedings{gilad2017algorand,
  title={Algorand: Scaling byzantine agreements for cryptocurrencies},
  author={Gilad, Yossi and Hemo, Rotem and Micali, Silvio and Vlachos, Georgios and Zeldovich, Nickolai},
  booktitle={Proceedings of the 26th symposium on operating systems principles},
  pages={51--68},
  year={2017}
}

\end{document}